\documentclass[twocolumn,aps,prl,reprint,nolongbibliography]{revtex4-2}

\usepackage{amsmath}
\usepackage{mathtools}

\usepackage{amssymb}
\usepackage{graphicx} 
\usepackage{verbatim} 
\usepackage{xcolor} 
\usepackage{subfigure} 
\usepackage{hyperref} 

\begin{document}
	
	\title{Lever rule violation and pressure imbalance in a driven granular system
	}
	\author{ Soumen Das, Anit Sane, Satyanu Bhadra}
	\affiliation{Department of Condensed Matter Physics and Materials Science, Tata Institute of Fundamental Research, Homi Bhabha Road, Mumbai 400-005, India}
 \author {Omer Granek, Yariv Kafri,  Dov Levine}
 \affiliation{Department of Physics, Technion-Israel Institute of Technology, Haifa, 3200003, Israel}
 	\author{ Shankar Ghosh }
	\affiliation{Department of Condensed Matter Physics and Materials Science, Tata Institute of Fundamental Research, Homi Bhabha Road, Mumbai 400-005, India}

	\date{\today}
	
	\begin{abstract}
    We study a monolayer of metal balls under periodic chiral driving in the horizontal plane. Energy dissipation occurs in this system via (i) inelastic collisions and (ii) frictional interaction with the substrate.  We show that below a density-dependent critical drive, the system phase separates into a fluid phase coexisting with a solid phase. Unlike ordinary coexistence, however, the system does not obey the lever rule, as the fluid-phase density depends on the overall particle density.  Additionally, the pressure is discontinuous across the fluid-solid interface, accompanied by a chiral edge current at the interface.
	\end{abstract}
	
	\maketitle

 A hallmark of phase coexistence is the so-called ``lever rule''~\cite{callen1991thermodynamics}: if the densities of two coexisting phases are $\phi_a$ and $\phi_b$, then the fraction $x$ of phase $a$ is given by $\phi=x \phi_a+(1-x)\phi_b$, where $\phi$ is the overall density. This relation stems from mass conservation, the convexity of the free energy, and the locality of the interactions. 
 
 Out of equilibrium, despite the lack of a free energy, the lever rule has been shown to obtain in a wide variety of systems. Examples include motility-induced phase-separation in active systems~\cite{tailleur2008statistical,fily_athermal_2012,redner_structure_2013,cates2015motility,solon_generalized_2018,solon_generalized_2018-1}, flocking~\cite{Solon2013,Solon2015b,agranov2024thermodynamically}, and driven granular materials~\cite{prevost2004nonequilibrium, PhysRevX.5.031025}. Despite the very different physics governing them, the lever rule is satisfied for all of these systems.

\begin{figure}[t]
		\centering
 		\includegraphics[width=1\linewidth]{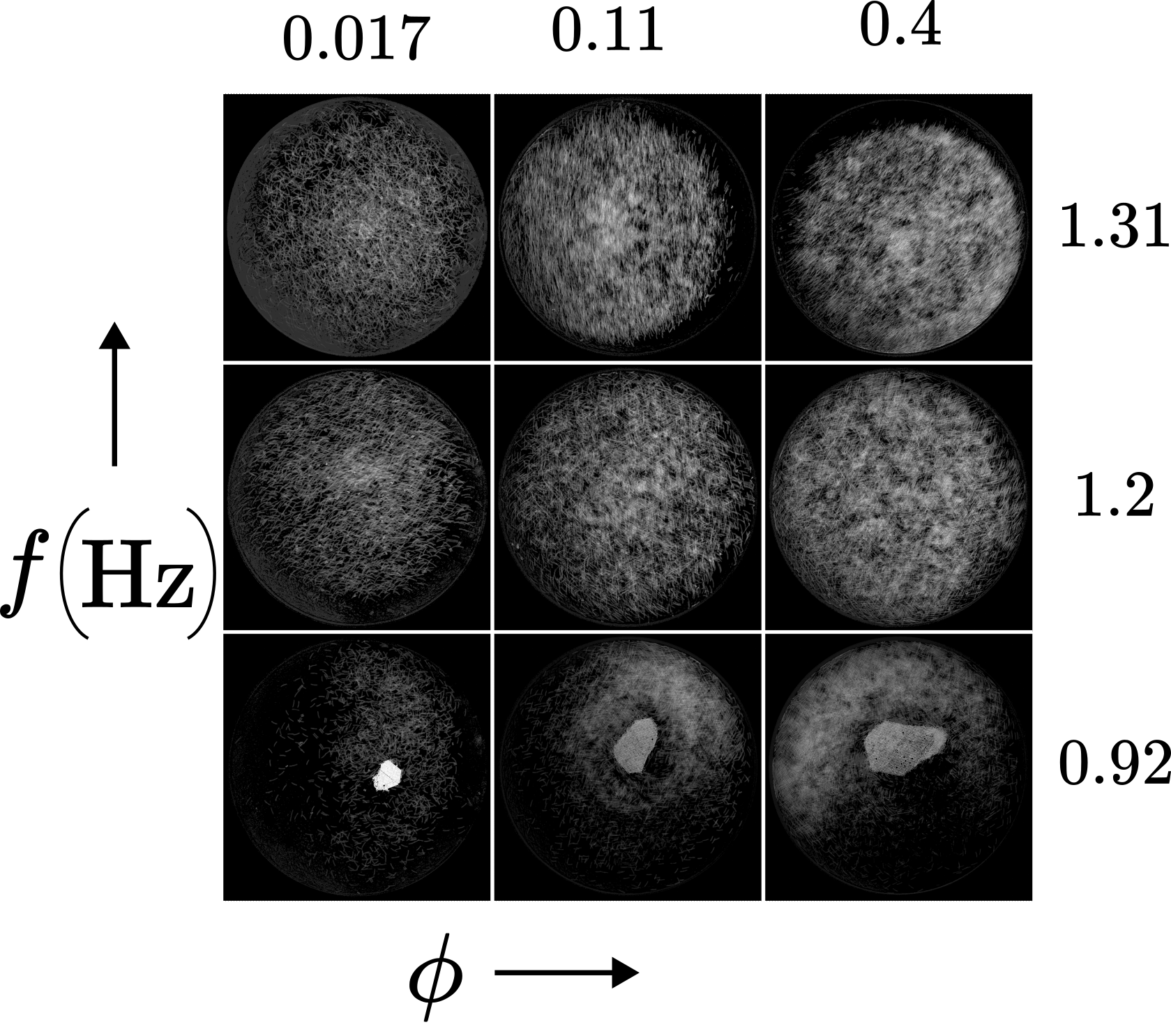}
		\caption{ Time averaged images of the chiral granular system in the co-moving frame, for some representative values of filling fraction $\phi$, driven at different frequencies $f$. For high $f$, the system is in an inhomogeneous swirling fluid phase, with all particles in motion. As $f$ is lowered beneath a $\phi$-dependent critical frequency  $f_c$, the system phase separates into a solid cluster surrounded by the fluid phase. For $f<f_m$ the system becomes stationary; this is not shown.}
		\label{fig:montage}
	\end{figure}

In this Letter, we present an experimental study of a driven granular system where spherical particles are driven periodically in a chiral fashion at a frequency $f$ while constrained to a plane  \footnote{Such a drive was employed to study size segregation as well as a fluid-solid transition at relatively high densities~\cite{aumaitre_granular_2003}.}. In a frequency range $f_m <f< f_c$, the system shows coexistence between a fluid phase and a solid phase; see Figure \ref{fig:montage}.  For $f<f_m$, the system is in a static phase with no particle motion since the drive is insufficient to overcome friction with the substrate. 
 For $f>f_c$, the system is in a homogeneous fluid phase. 
 
\begin{figure}
    \centering
    \includegraphics[width=1\linewidth]{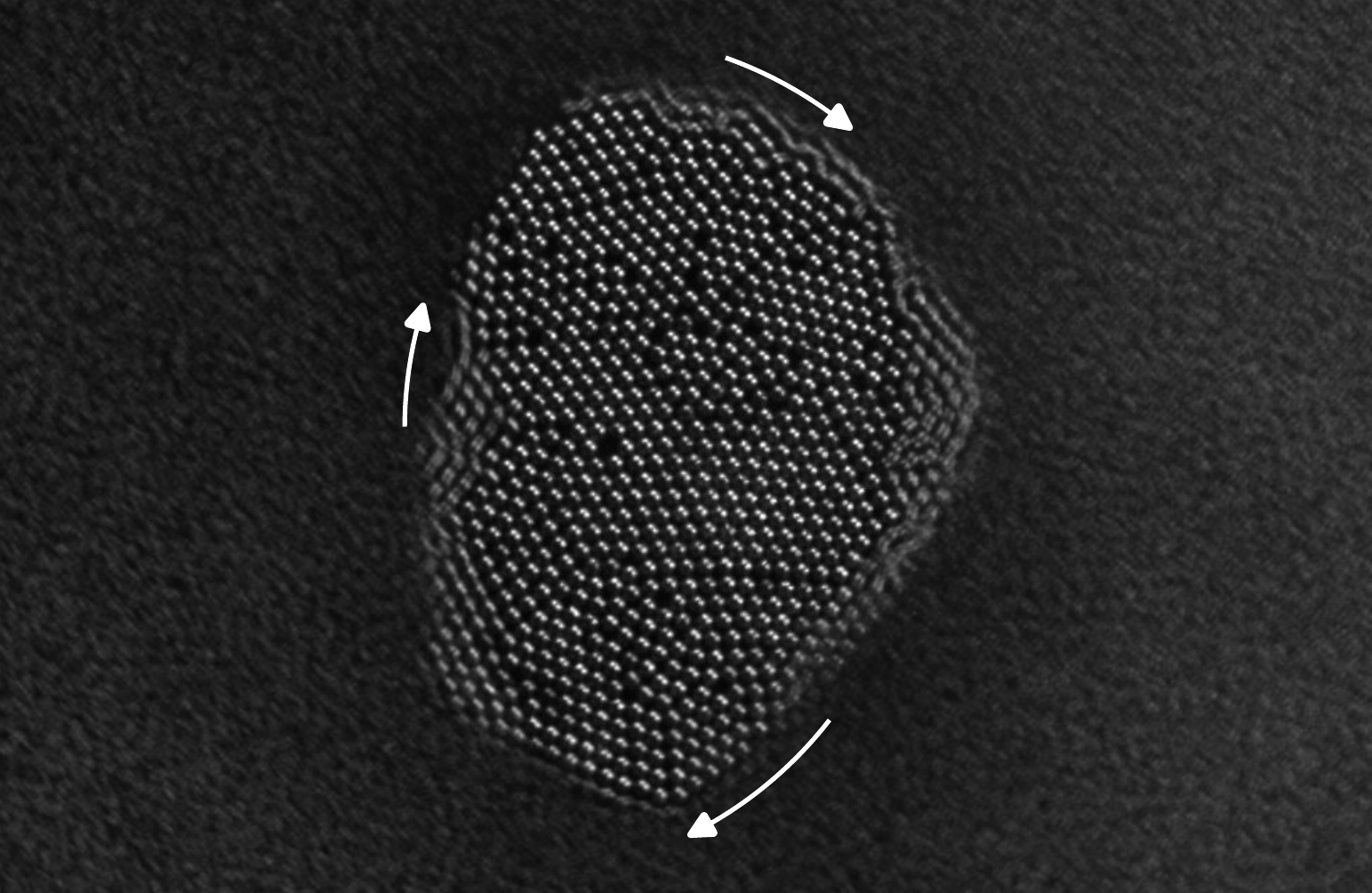}
    \caption{Time-averaged picture of the coexistence of an immobile solid phase inside a fluid phase for $\phi\approx0.04$ and $f=0.8$Hz. The fluid phase is blurred by the time average. At the solid-fluid interface, there is a chiral edge current of particles in the direction of the white arrows and opposite to the direction of the drive. The particles comprising the current are partially blurred due to their motion. The images are averaged over approximately one period.}
    \label{fig:chiral current}
\end{figure}

Among its rich and unusual behavior, we observe:
\begin{itemize}
    \item Violation of the lever rule in the coexistence region, with the fluid-phase density depending on the filling fraction at constant drive.
     \item In the coexistence region, the pressures in the solid and fluid phases are unequal.
    \item The existence of a chiral edge current around the solid cluster, as seen in Figure \ref{fig:chiral current}.
   
\end{itemize}
As we will argue later, the system bears some similarities to chiral active fluids, which have received much attention in recent years~\cite{VanZuiden2016,Markovich2019,Han2021, Tan2022,Shankar2022,Kalz2023,Mecke2024}, but shows significant differences as well. In particular, hydrodynamic theories of active systems predict that the lever rule is obeyed; see, for example, References ~\cite{wittkowski_scalar_2014,Speck2014,cates2015motility,solon_flocking_2015,solon_generalized_2018,agranov2024thermodynamically}.  This remains true when the chiral drive is taken into account, as discussed in Section S4 of the Supplemental Material. Our observation that the lever rule is violated appears to imply that the system can not be captured by a local field theory.

	     \begin{figure}[t]
        \includegraphics[width=0.95\linewidth]{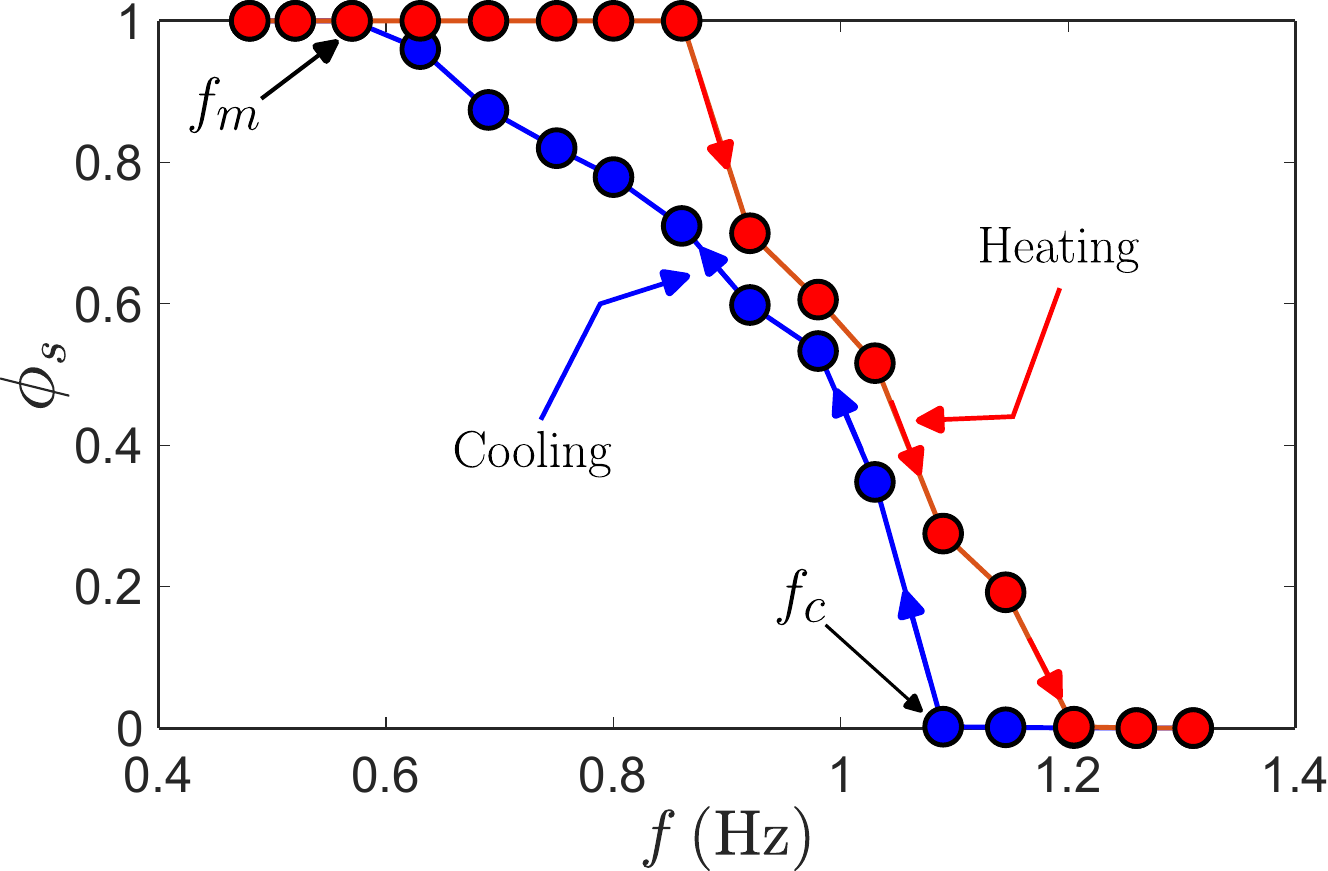}
		\caption{``Cooling'' (blue) and ``heating'' (red) trajectories for $\phi=0.016$, showing the fraction of particles $\phi_s$ in the solid cluster. The time direction of the protocol is indicated by the arrows.  At high frequencies, the entire system is in the swirling fluid phase, and $\phi_s = 0$.
 There is evident hysteresis in a full cycle of cooling and heating.}
	\label{fig:hysteresis}
    \end{figure}

	
	In the experiment, $N$  stainless steel balls of diameter $2a=0.8 \,\rm{mm} $   are confined in a cylindrical container of diameter $ D=150 \,\rm{mm} $ and height $ 1 \,\mathrm{mm} $ with an anodized aluminum substrate. We will express the filling fraction, or total density of particles, by the area fraction $\phi=N (2a/D)^2$. This assembly, along with a camera, is made to perform an orientation-preserving horizontal circular motion such that each point on the plate moves in a circle of the same radius $r$~\cite{aumaitre2001segregation,kumar2015granular,kumar2014spreading}.       The experiments are performed for a constant drive amplitude $r=12 ~\rm{mm} $ and varying frequencies $ f $.

	The friction force exerted at the point of contact between a static sphere and the moving plate produces a torque that induces a rolling motion of the balls when the force acting on a ball exceeds the force of static friction. A detailed description of the mechanics associated with this problem can be found in Ref.~\cite{das2022state}. An isolated sphere on an orbitally moving platform performs an approximate swirling motion~\cite{das2022state,scherer2000swirling,aumaitre2001segregation,aumaitre_granular_2003}; such trajectories are randomized by collisions with the boundary and other particles.
		 The velocity distribution is unimodal with an exponential tail 
        and a mean which scales linearly with the  frequency of the drive, i.e., $\bar{v} \propto f$ (see SM Sec.~S1).


In what follows, we map out the phase diagram, which comprises three phases. For $f<f_m$, the drive is insufficient to overcome static friction with the substrate, and particles are stationary at all $\phi$. For $f_m<f<f_c(\phi)$, the system  is phase separated into a crystal-like region of area fraction $\phi_s$ and a swirling fluid phase. This swirling phase consists of two regions of roughly homogeneous densities: a dense region and a very dilute one. As seen in Fig.~\ref{fig:montage}, the very dilute region seems to result from a shadowing effect of the crystal. Thus, in what follows, the area fraction $\phi_f$ of the swirling fluid phase is always taken to be that of the high-density region. Finally, for $f>f_c(\phi)$ the system is a homogeneous fluid phase.

For a given filling fraction $\phi$, we map out the system behavior with a `cooling protocol', where we start from a random initial state at a frequency $f>f_c(\phi)$ so that the entire system is in a homogeneous swirling fluid phase. Specifically, we take $f=1.5~{\rm Hz}$. We then slowly decrease the frequency in steps of $0.03~{\rm Hz}$, holding the system at the new frequency for about $600~{\rm s}$.  A typical run is indicated by the blue curve in Figure \ref{fig:hysteresis}.  As $f$ is decreased, the homogeneous swirling fluid phase persists until a density-dependent critical frequency $f_c(\phi)$, at which a stationary solid cluster forms and coexists with the fluid. This is reminiscent of the motility-induced phase separation (MIPS) in active matter~\cite{fily_athermal_2012,redner2013reentrant,redner2013structure,cates_when_2013,cates2015motility}, where a condensed liquid phase develops and coexists with a gaseous phase. Here, the fraction of the system in the solid cluster rises sharply as we cross the transition, as seen in Fig.~\ref{fig:hysteresis}. As seen in Fig.~\ref{fig:hysteresis}, as $f$ is decreased further, the cluster grows continuously  until  $f=f_m(\phi)$, at which point all particles are in stationary solid clusters.  Figure~\ref{fig:montage}  shows time-averaged images of representative configurations characterized by $\phi$ and $f$.  
  
  We note that the protocol going between the fluid and static phases displays hysteretic behavior:  a `heating protocol,' where we start from a distribution of static balls, is inequivalent to the cooling protocol; see Fig.~\ref{fig:hysteresis}. The hysteresis seems to result from friction with the substrate, whereby static particles will only begin to move if the frequency is increased to overcome static friction.

\begin{figure}[t]
	\centering
        \includegraphics[width=1.\linewidth]{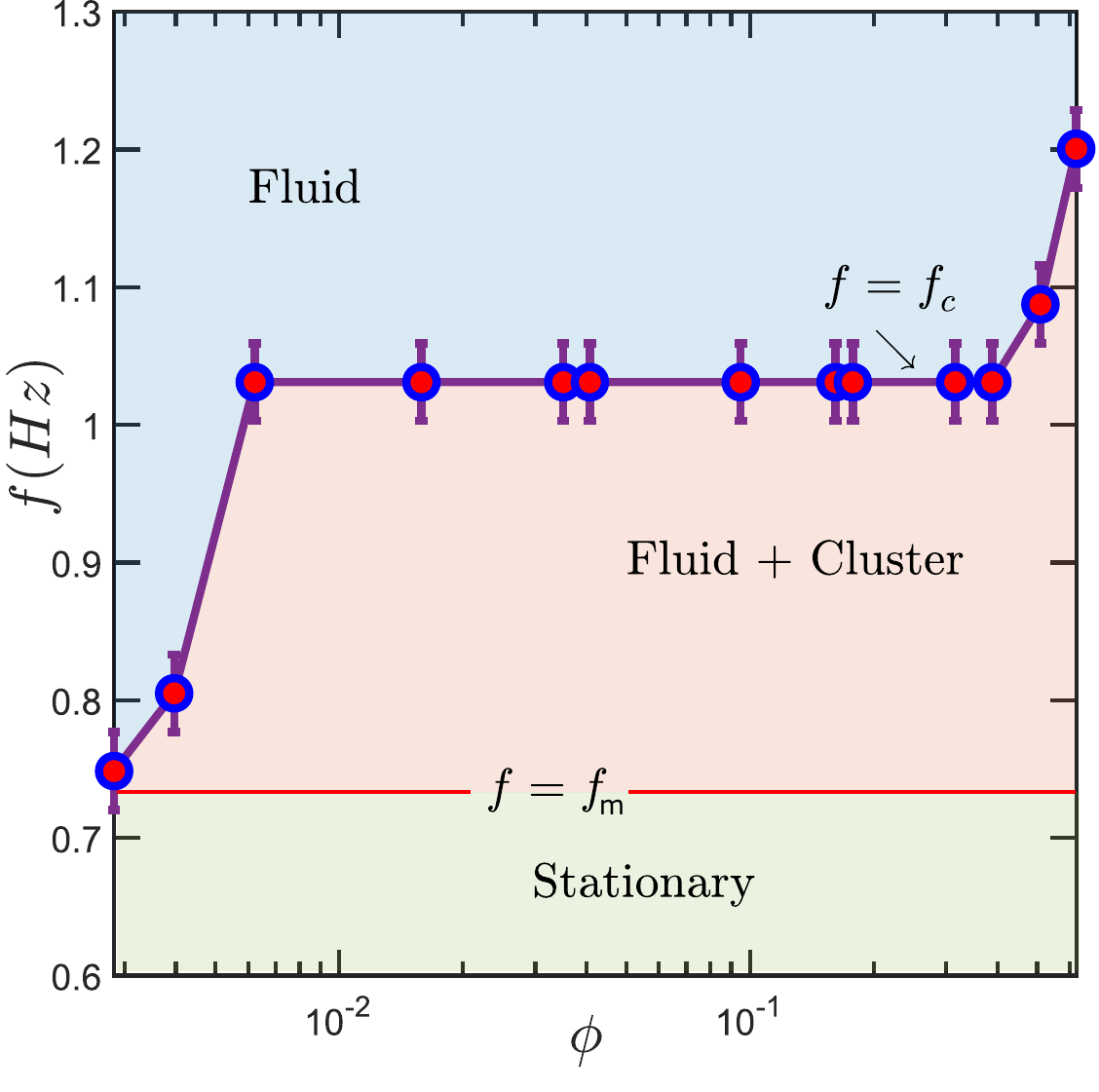}
		\caption{ Phase diagram showing the coexistence region of the solid and fluid phases.  Data points indicate the critical value of the frequency at which cluster formation initiates, as a function of  the filling fractions $\phi$, in the cooling protocol.  Beneath $f=f_m$ all the particles are static. }
	\label{fig:phase}
    \end{figure}

 In the coexistence region $f_m < f < f_c(\phi)$, the solid phase consists of crystalline clusters separated by grain boundaries, reminiscent of the mosaic phase discussed in Ref.~\cite{caporusso_motility-induced_2020}.    
 The experimentally measured phase diagram is shown in Fig.~\ref{fig:phase}, in which the values of $f_c$ are taken in the cooling protocol.

	\begin{figure}[t]
 		\centering
         \includegraphics[width=1\linewidth]{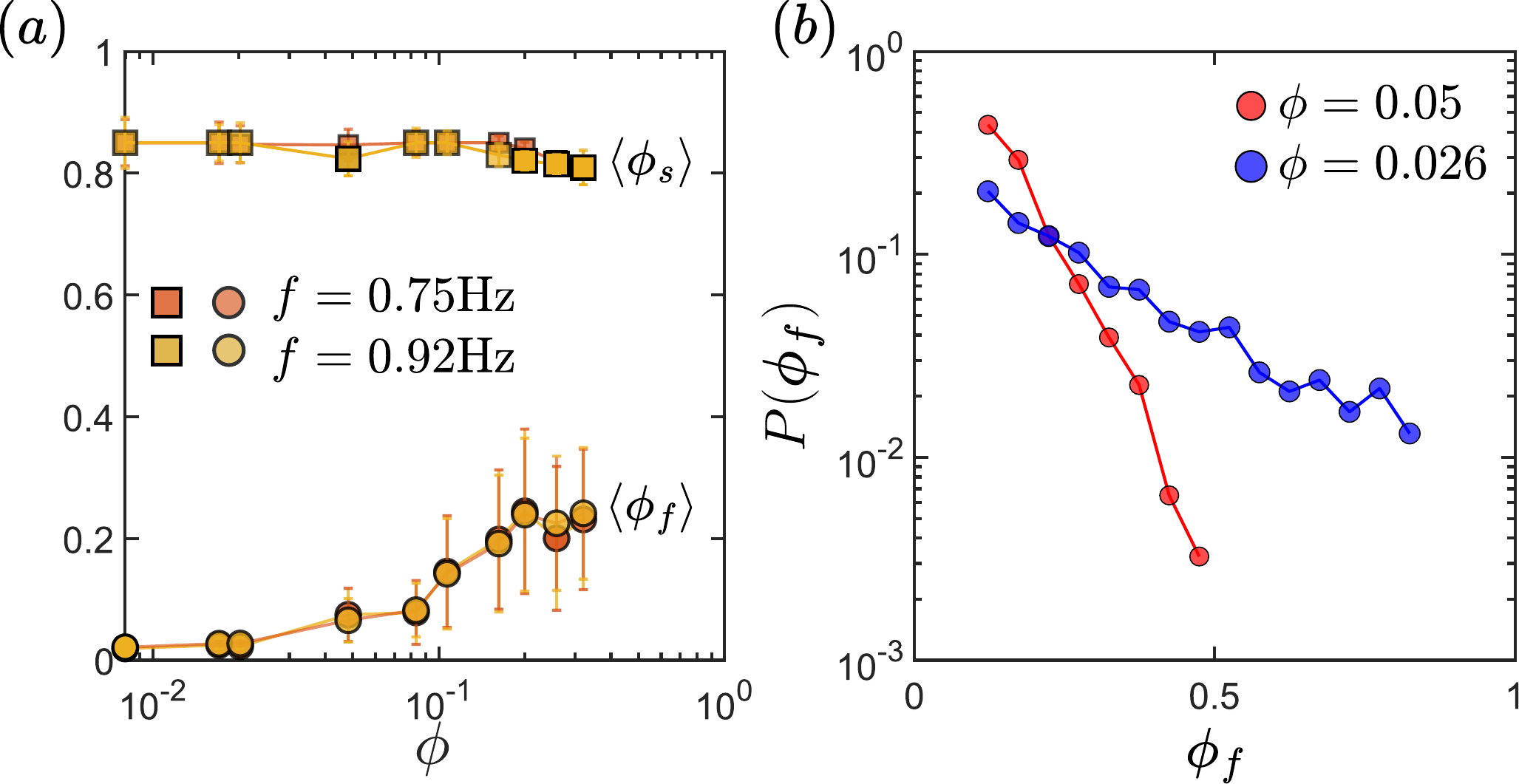}
 		\caption{(a) Density of solid and fluid phases as a function of the overall filling fraction $\phi$. The density of the solid phase is essentially independent of $\phi$, while the density of the fluid phase increases linearly with $\phi$. This shows that the density of the fluid phase is not only a function of the drive frequency, indicating a violation of the lever rule.   (b) The probability distribution function of the liquid phase density at two values of  $\phi$ for $f=0.92$ Hz; see SM Sec.~S1 for details.}
 		\label{fig:area_fraction}
 	\end{figure}

Perhaps the most striking behavior shown by the system is the violation of the ``lever rule'' of phase coexistence~\cite{callen1991thermodynamics}.	 In the standard picture of coexistence, for given values of the system control parameters (here, the driving frequency $f$), the system separates into two phases, one dense (solid) and one dilute (fluid).  These two phases have well-defined characteristic densities $\phi_s$ and $\phi_f$, which are obtained by minimizing a local free-energy functional, and depend only on the control parameters.  If we increase (decrease) the overall density of the system, a greater (lesser) fraction of the system converts to the dense phase, but the densities $\phi_s$ and $\phi_f$ remain unchanged. This is not merely a manifestation of thermodynamic equilibrium -  it holds even in classes of active matter; see, e.g., Refs.~\cite{cates2015motility,Solon2013,Solon2015b,solon_generalized_2018,agranov2024thermodynamically}.  As we show in the SM Sec.~S4, this continues to hold in a simple local chiral active field theory.

Surprisingly, our system violates the lever rule.  As seen in Fig.~\ref{fig:area_fraction}a, the density of the solid phase $\phi_s$ is indeed nearly constant. However, the density of the fluid phase is seen to increase with $\phi$. This means that as we add more particles to the system, a disproportionate number go into the fluid phase, irrespective of $\phi$. Hence, the fluid phase is not characterized by a well-defined density $\phi_f$ that depends on $f$ alone.    We are not aware of any other system, theoretical or experimental, that demonstrates this behavior.
		
Next, we consider pressure measurement in the two phases. In recent years, it has become clear that for non-equilibrium systems that do not conserve momentum, the mechanical pressure displays many anomalous features~\cite{solon2015pressure,fily_mechanical_2017,solon_generalized_2018,solon_generalized_2018-1,Granek2024}. In particular, it is expected on general grounds that two coexisting phases will have different pressures~\cite{fily_mechanical_2017}. However, this theoretical claim has never been verified in experiments so far.

\begin{figure}[t]
		\centering
        \includegraphics[width=.9\linewidth]{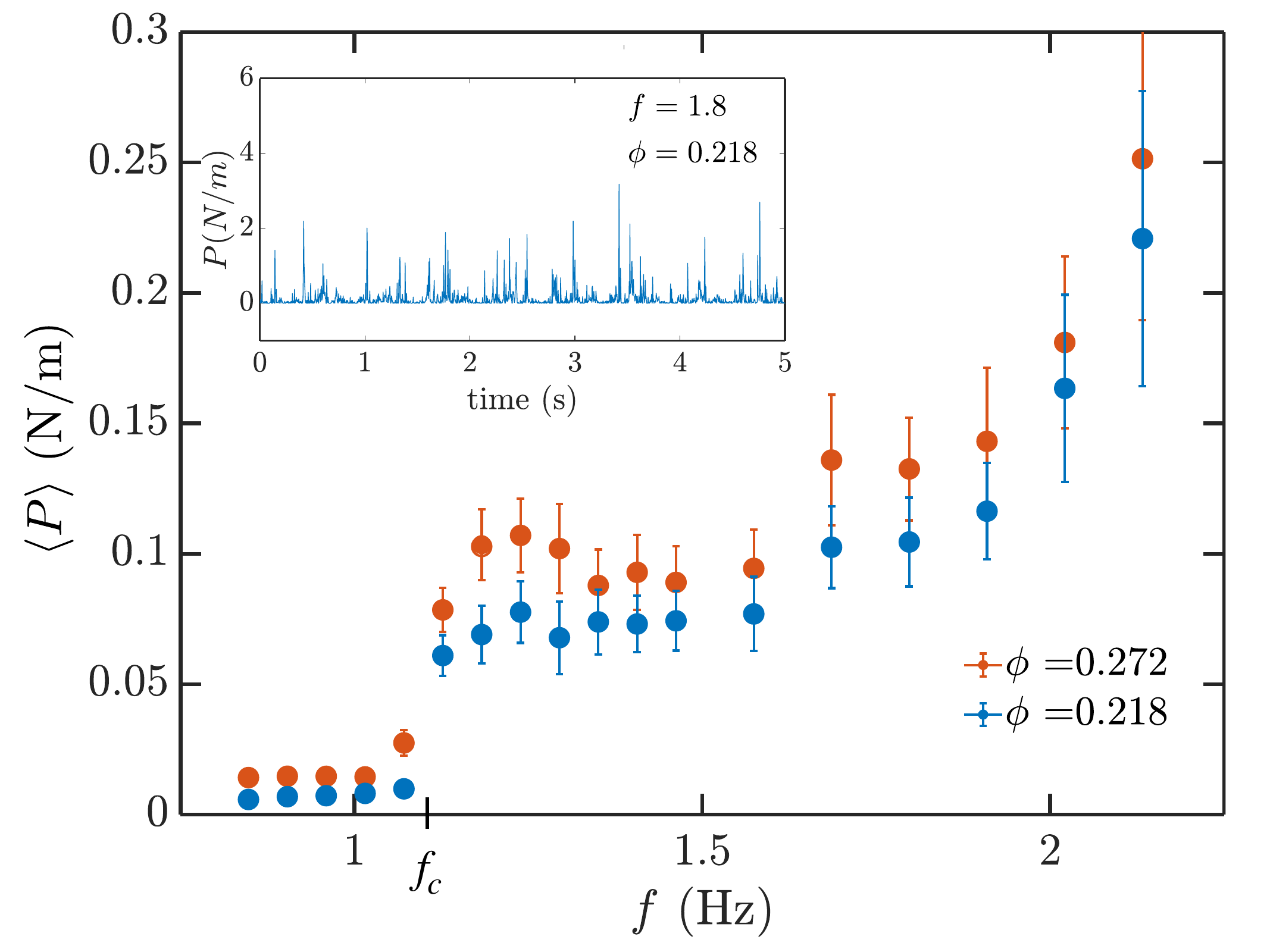}
		\caption {Variation of the average pressure $\langle P \rangle$ as as a function of the drive frequency for two representative  values of packing fraction $\phi$.  The inset shows a typical time scan of the instantaneous pressure $P$. The average pressure, $\langle P \rangle$, jumps discontinuously at the critical frequency $f=f_c$.  For $f = f_c^-$, just below $f_c$, the sensor is inside of a crystal, so that it measures the pressure inside the solid phase,  For $f = f_c^+$, just above $f_c$, it measures the pressure in the fluid phase.  The jump is the difference in the pressure in the two phases near criticality.
  } 
		
		\label{fig:pressure1}
	\end{figure}

To address this issue, we introduced a deformable circular silicone sensor into the system to measure the mechanical force (see SM Sec.~S2 for details of the measurement technique). The sensor deforms, measuring the magnitude of its collisions with particles, and providing a time series $P(t)$, as seen in the inset to Figure \ref{fig:pressure1}.  This time series is then integrated over time to give an average pressure $\langle P \rangle$,
\begin{equation}
\langle P \rangle \;= \;\frac{1}{T} \int_0^T P(t)dt
\end{equation}

In the fluid phase, the sensor experiences collisions with individual particles, and the time series is a random sequence of spikes whose height and frequency both scale as the mean particle velocity $\bar v$.  Since $\bar v \propto f$, we expect $\langle P \rangle \sim f^2$ in the fluid, which is confirmed in Figure \ref{fig:pressure1} and the SM Sec.~S3. In the coexistence region, the sensor provides a nucleation site around which a solid cluster grows, leading to the measurement of the pressure inside the solid. In this case, the frictional interaction of the solid particles and the substrate screens the sensor from collisions of fluid particles with the boundary of the solid cluster.

Figure \ref{fig:pressure1}  shows the variation of  $\langle P \rangle$  as a function of the driving frequency $f$ for several values of filling fraction $\phi$. 
As $f$ decreases from a value $f>f_c$, $\langle P \rangle$ decreases monotonically and jumps at $f=f_c$.    As $f$ is decreased further, $\langle P \rangle$ continues to decrease monotonically. The fluid phase immediately above and below $f=f_c$ remains essentially unchanged.  Thus, although we can not measure it directly, since the sensor is inside the solid cluster, we infer that the pressure of the fluid phase in the coexistence state, just below $f_c$, is identical to that in the homogeneous phase just above $f_c$: $\langle P_{fluid}(f_c^+)\rangle \approx \langle P_{fluid}(f_c^-)\rangle$. From this we infer that the pressure in the cluster is different from that in the coexisting fluid, with a pressure difference proportional to the jump in  $\langle P \rangle$ at $f_c$. This jump implies that, unlike equilibrium systems, the pressure of the chiral granular system is not a good state variable with a well-defined equation of state. That pressure may not be a state variable in non-equilibrium systems has been established for active matter~\cite{solon2015pressure,junot2017active}, where the nature of the confining walls, for example, may influence the pressure due to the injection of momentum from the substrate in their vicinity~\cite{fily_mechanical_2017}. This lead in some cases to coexistence between two phases with different pressures. To our knowledge, this is the first time the phenomenon has been measured experimentally.


The last feature we wish to discuss is the existence of a chiral current around the solid cluster.  Figure~\ref{fig:chiral current} shows a time-averaged image of the cluster coexisting with the swirling fluid phase. Since the camera is co-moving, the particles in the cluster are sharp, as they are static in this frame of reference.  The particles in the surrounding fluid phase are grey and blurred since they have moved during the exposure.  Around the periphery of the cluster, there is a current of particles whose direction of motion is indicated by the white arrows.  For reference, the motion of the substrate is in the opposite sense.   
Steady chiral edge currents have been widely observed in experiments on chiral non-equilibrium systems under confinement~\cite{Mecke2024}, such as nematic cell monolayers~\cite{Yamauchi2020,Yashunsky2022}, chiral bacterial suspensions~\cite{Beppu2021} and active granular spinners~\cite{Yang2020,Petroff2023}. Such currents were reproduced by simplified models~\cite{VanZuiden2016,Dasbiswas2018,Caprini2019,Yang2021_2}. Moreover, chiral edge currents in phase-separating interfaces have been observed in experiments on spinning colloids~\cite{Soni2019,Massana-Cid2021,Liebchen2022} and in simulations of chiral active particles~\cite{Reichhardt2019,Adorjani2024,Yuan2024} and chiral Lennard-Jones particles~\cite{Caporusso2024}. 

The existence of a chiral edge current in the phase-separating interface
can be demonstrated based on symmetry considerations as follows:

The current
$\mathbf{J}({\bf r},t)$ is defined via the mass-conservation equation
$\partial_{t}\rho=-\nabla\cdot{\bf J}$, where $\rho({\bf r},t)$
is the mass density. We describe the dynamics in the co-rotating frame
and make two observations. First, the system's bulk adheres to translation
invariance: each point rotates about its own center, and there is no rigid-body rotation about a common center.
Second, the system is isotropic and chiral, i.e., chiral symmetry
is explicitly broken by the driving frequency ${\bf \Omega}=2\pi f\hat{{\bf z}}$.
Therefore, the only vectors on which $\bf{J}$ can depend are
${\bf \nabla}=\hat{{\bf x}}\partial_{x}+\hat{{\bf y}}\partial_{y}$
and ${\bf \nabla}_{\perp}=-\hat{{\bf x}}\partial_{y}+\hat{{\bf y}}\partial_{x}$.
Hence, ${\bf J}$ admits the general decomposition
\begin{align}
{\bf J} & ={\bf J}_{\parallel}+{\bf J}_{\perp},\\
{\bf J}_{\alpha}({\bf r},t) & =-M_{\alpha}({\bf r},t)\bm{\nabla}_{\alpha}\mu_{\alpha}({\bf r},t),
\end{align}
where $\alpha\in\{\perp,\parallel\}$ and $M_{\alpha}$ and $\mu_{\alpha}$
are unknown scalars. Given an interface of finite width connecting
two homogeneous phases, $\mu_{\alpha}({\bf r},t)$ jumps abruptly
across the interface between two constant values. Therefore, $\bm{\nabla}\mu_{\parallel}=\hat{{\bf n}}\partial_{n}\mu_{\parallel}$
and $\bm{\nabla}_{\perp}\mu_{\perp}=\hat{{\bf t}}\partial_{n}\mu_{\perp}$,
where $\hat{{\bf n}}$ and $\hat{{\bf t}}$ respectively are the unit
vectors normal and tangential to the interface and $\partial_{n}\equiv\hat{{\bf n}}\cdot\bm{\nabla}$
is the normal derivative. Thus, ${\bf J}_{\parallel}\propto\hat{{\bf n}}$
and ${\bf J}_{\perp}\propto\hat{{\bf t}}$. In the steady-state, spatial
isotropy implies that ${\bf J}_{\parallel}=0$. It follows that a chiral current ${\bf J}\propto\hat{{\bf t}}$
exists solely inside the interface. The emergence of the current can
be seen as a manifestation of the Hall effect, where a gradient in
a scalar quantity is coupled to a current transverse to the gradient~\cite{Hurd1972,Avron1995,Larralde1997,Avron1998,Banerjee2017,Shankar2022,Hargus2021}.
In contrast with standard Hall systems, here, the gradient is confined
to the cluster's edge, providing a chiral edge current.

In our study, frictional inelastic balls are driven in a chiral fashion.  This chiral granular system exhibits condensation from a homogeneous fluid phase to a solid phase coexisting with the fluid.  Although this is, at first glance, similar to the MIPS transition in conventional active matter such as Active Brownian particles, significant differences are seen.  This includes the violation of the lever rule of coexisting phases and a difference in pressure between the coexisting phases.  It is of interest to understand how essential the chirality of driving is, which we will explore in a future publication. 

\emph{Acknowledgements:}  We would like to thank Julien Tailleur for interesting discussions.  DL thanks the Israel Science Foundation for grant 2083/23. YK and OG acknowledge financial support from ISF (2038/21) and NSF/BSF (2022605). OG also acknowledges support from the
Adams Fellowship Program of the Israeli Academy of Science and Humanities. SG acknowledges the support of the Department of Atomic Energy, Government of India, under Project No. 12-R \&DTFR-5.10-0100.

%

\end{document}